\documentclass[oldversion]{aa}

\usepackage{epsfig}
\usepackage{graphics}
\usepackage{float}
\usepackage{amsmath}
\usepackage{multirow}
\usepackage{longtable}
\usepackage{rotate}
\usepackage{array}
\usepackage{subfigure}
\def\kms{\ifmmode{\rm km\,s^{-1}}\else\hbox{$\rm km\,s^{-1}$}\fi}

\setlongtables

\begin{document}

\title{Binary orbits from combined astrometric and spectroscopic data}

\author{L.B.Lucy}
\offprints{L.B.Lucy}
\institute{Astrophysics Group, Blackett Laboratory, Imperial College 
London, Prince Consort Road, London SW7 2AZ, UK}
\date{Received ; Accepted }

\abstract{An efficient Bayesian technique for estimation problems in 
fundamental stellar
astronomy is tested on simulated data for a binary observed both
astrometrically and spectroscopically. Posterior distributions are computed
for the components' masses and for the binary's parallax. One thousand 
independent repetitions of the simulation demonstrate that the 1- and
2-$\!\sigma$ credibility intervals for these fundamental quantities have 
close to the correct coverage fractions. In addition, the simulations allow 
the investigation
of the statistical properties of a Bayesian goodness-of-fit criterion
and of the corresponding $p$-value. The criterion has closely similar
properties to the traditional $\chi^{2}$ test for 
minimum-$\chi^{2}$ solutions. 
}

\authorrunning{Lucy}
\titlerunning{Binary orbits}
\maketitle

\section{Introduction}

In fundamental stellar astronomy, all statistical estimation problems 
involve mathematical models with both linear and 
non-linear parameters - the so-called hybrid problems. 
The linear parameters determine scale and location; the non-linear
parameters appear as arguments in dimensionless functions of time.

The presence of linearities suggests that more efficient 
estimation techniques exist than when all parameters are treated as
non-linear (Wright \& Howard 2009; Catanzarite 2010). However, some attempts 
to achieve this lead to 
significantly underestimated error bars (Eastman et al. 2013 ; 
Lucy 2014, L14a). 
This poses the
challenge of developing a technique that achieves the computational efficiency 
allowed by linearity while still giving confidence or credibility intervals
with correct coverage - so that, for example, 1-$\sigma$ error bars contain
the correct answer with probability 0.683. A solution to this challenge
is presented in Lucy (2014, L14b) where a grid search in the space defined by
the non-linear parameters is combined with Monte Carlo sampling of
the space defined by the linear parameters. In L14b, this technique is
applied to simulated observations of a visual binary and coverage experiments
confirm that $1-$ and 2-$\!\sigma$ error bars enclose the
exact values with close to the frequencies expected for normally-distributed 
errors.

In this paper, a significantly harder problem is posed, that of 
analysing a binary with both astrometric and spectroscopic data. Such data
could be analysed separately, but this is sub-optimal since information
concerning several orbital elements is present in both data sets. 
Accordingly,
the aim here is to obtain the posterior distribition over the entire
parameter space using both data sets and to test if the
derived error bars are trustworthy, an essential requirement for fundamental 
data on stellar masses and luminosities.

The posed problem aims at demonstrating proof-of-concept in the treatment
of hybrid problems and to provide a template for the many such problems
in statistical astronomy. To this end, the code developed for this
investigation is freely available.

Although here a technical exercise, the simultaneous analysis of astrometric
and spectroscopic data is of practical importance in the era of adaptive
optics (AO) and speckle interferometry. As emphasized by Mason et al. (1999),
the ability to resolve binary stars at or near the diffraction limit
results in a powerful synergy between short-period visual and long-period
spectroscopic binaries, leading to stellar masses and improved 
mass-luminosity relations.  

\section{Synthetic orbits}

The physical model comprises an isolated pair of stars undergoing
Keplerian motion due to their mutual gravitational attraction. This binary
is observed astrometrically and spectroscopically, yielding two independent
data sets $D_{a}$ and $D_{s}$, respectively. 
To analyse these data sets,
the mathematical models predicting the components'
relative motion on the sky as well as their radial velocity variations
are used simultaneously to derive the posterior distribution  
over parameter space.
\subsection{Orbital elements} 
For the astrometric orbit, L14a is followed closely with 
regard both to notation and the 
creation of synthetic data.

The motion on the sky of the secondary relative to the primary is 
conventionally
parameterized by the  Campbell elements 
$P,e,T,a,{\rm i},\omega,\Omega$. 
Here $P$ is the period, $e$ is the eccentricity,
$T$ is a time of periastron passage, ${\rm i}$ is the inclination, 
$\omega$ is the longitude of 
periastron, and $\Omega$ is the position angle of the ascending node.
However, from the standpoint of computational economy, many investigators
- references in L14a, Sect.2.1 - prefer the Thiele-Innes elements.
Thus, the Campbell parameter vector $\theta = (\phi,\vartheta)$, where
$\phi = (P,e,\tau)$ and $\vartheta = (a,{\rm i},\omega,\Omega)$, is 
replaced by the Thiele-Innes vector $(\phi,\psi)$, where
the components of the vector $\psi$ are the Thiele-Innes constants
$A,B,F,G$.   
(Note that in the $\phi$ vector,
$T$ has been replaced by $\tau = T/P$ which by definition $\in (0,1)$.)

The spectroscopic orbits of the components introduce three additional
parameters, the systemic velocity $\gamma$ and the semi-amplitudes $K_{1,2}$.
The predicted radial velocities are then 
\begin{equation}
 v_{1,2}(t) = \gamma + K_{1,2} \: [ cos(\nu + \omega_{1,2}) 
                                + e \: cos \: \omega_{1,2} ]
\end{equation}
where $\nu(t)$ is the true anomaly, $\omega_{2} = \omega$ and 
$\omega_{1} = \omega + \pi$.

Note that $v_{2} - v_{1} = \dot{z}$, where $z$, the companion's  
displacement perpendicular to the sky, is given by the Thiele-Innes
constants $C$ and $H$. This is a useful check on coding.
 
With the inclusion of spectroscopic data, the combined
data sets allow inferences about the 10-dimensional vector 
\begin{equation}
 \Theta = (\phi, \psi, \lambda)
\end{equation}
where $\lambda = (\gamma, K_{1}, K_{2})$.

In a Bayesian analysis, the task is to compute the posterior probability
density in $\Theta$-space given $D_{a}$ and $D_{s}$.
\subsection{Model binary} 
The adopted  model binary has the following Campbell elements:
\begin{eqnarray}
  P = 10y  \;\;\; \tau = 0.4 \;\;\; e = 0.6  \;\;\; a = 0.3\arcsec  
                                                   \nonumber    \\
  {\rm i} = 70\degr   \;\;\;    \omega  = 250\degr    \;\;\;  
                                              \Omega  = 120\degr
\end{eqnarray}
With this $a$, the binary would be unresolved in seeing-
broadened images but should be resolved in images approaching
diffraction limits.

If we now take the parallax $\varpi = 0.05\arcsec$, the total mass
$ = 2.16 {\cal M}_{\sun}$. With mass ratio $q = 0.7$, the component masses
are ${\cal M}_{1} = 1.27$ and ${\cal M}_{2} = 0.889 {\cal M}_{\sun}$. The 
resulting semi-amplitudes are $K_{1} = 8.64$ and $K_{2} = 12.35$km s$^{-1}$, and
we take $\gamma = 0.0$km s$^{-1}$.
\subsection{Observing campaigns} 
An astrometric observing campaign is simulated by creating 
measured Cartesian sky coordinates $(\widetilde{x}_{n},\widetilde{y}_{n})$ with weights
$w_{n}^{a} = 1/\sigma_{a,n}^{2}$ for both coordinates. One observation in 
observing
seasons of length $0.3$y is created randomly for 10 successive years. We take 
$\sigma_{a,n} =  0.05\arcsec$ for all $n$.

In the above, we assume equal precision for each coordinate {\em and}
uncorrelated errors. These assumptions are well justified if the AO or
speckle image 
reconstructions give circularly symmetric stellar profiles. If necessary,
the technique can be generalized to treat unequal and correlated errors
(Sects. A.1, C.1).
   
A spectroscopic observing campaign is simulated by creating measured radial
velocities $\widetilde{v}_{1n},\widetilde{v}_{2n}$ at random times in 10 successive
observing seasons. The observations have weights 
$w_{1n}^{s} =  1/\sigma_{s,n}^{2}$ and  $w_{2n}^{s} =  0.5/\sigma_{s,n}^{2}$.
We take  $\sigma_{s,n} = 0.5$km s$^{-1}$ for all $n$. 

All simulated measurement errors are normally-distributed.
\section{Conditional probabilities}
In order to benefit from the hybrid character of the problems arising in 
orbit estimation, the chain rule for conditional probabilities is used
to factorize multi-dimensional posterior distributions $\Lambda$ in such a 
way that
linear and non-linear parameters are separated. This facilitates the
construction of efficient hybrid numerical schemes that combine grid
scanning with Monte Carlo sampling. 
\subsection{Approach}
Consider a problem with two scalar parameters, $\alpha$ and $\beta$, and
suppose the model is non-linear in $\alpha$ and linear in $\beta$. Applying
the chain rule, we can write the posterior density as 
\begin{equation} 
 \Lambda(\alpha,\beta)  = Pr(\alpha) \: Pr(\beta|\alpha)
\end{equation}
where
\begin{equation} 
  Pr(\alpha) =  \int \Lambda(\alpha,\beta) \: d \beta
\end{equation}
$Pr(\alpha)$ is thus the projection of $\Lambda(\alpha,\beta)$
onto the $\alpha$ axis.

The 1-D function $Pr(\alpha)$ can be approximated by the discrete values
$Pr(\alpha_{i})$, where the $\alpha_{i}$ are the mid-points of a uniform
grid with steps $\Delta \alpha$. In contrast, because of linearity,
${\cal N}_{i\ell}$ values $\beta_{i \ell}$ can readily be derived that randomly
sample $Pr(\beta| \alpha_{i})$. Combining these approaches, we derive the
following approximation for the posterior distribution 
\begin{equation} 
 \Lambda(\alpha,\beta)  = \sum_{i\ell} \Delta \alpha  \:  Pr(\alpha_{i}) 
 \times {\cal N}_{i \ell}^{-1} \: \delta(\beta - \beta_{i \ell})
\end{equation}
With this approximation, all the quantities we wish to infer from the posterior
distribution became weighted summations over the points 
$(\alpha_{i}, \beta_{i \ell})$,
and these summations converge to exact values as $\Delta \alpha \rightarrow 0$
and ${\cal N}_{i \ell} \rightarrow \infty$. Arbitrary accuracy can therefore
be achieved.
\subsection{Astrometry only}
If we only have astrometric data, $\Lambda$ is a
function of seven parameters. With the Thiele-Innes parameterization,
the parameter vector is $(\phi,\psi)$, and the 
mathematical model is linear in the four $\psi$ parameters and non-linear in
the three $\phi$ parameters. 

Following the 2-D example of Sect.3.1, we apply the chain rule to obtain
\begin{equation} 
 \Lambda(\phi,\psi| D_{a})  = Pr(\phi|D_{a}) \: Pr(\psi|\phi,D_{a})
\end{equation}
where
\begin{equation} 
  Pr(\phi|D_{a}) =  \int \Lambda \: d\psi
\end{equation}
Here $Pr(\phi|D_{a})$ is the projection of the 7-D posterior distribution
$\Lambda(\phi,\psi|D_{a})$ onto the 3-D $\phi$-space. The second factor
$Pr(\psi|\phi,D_{a})$ then specifies how this projected or summed 
probability is to be distributed in $\psi$-space. 
\subsection{Astrometry and spectroscopy}
With spectroscopic data included, $\Lambda$ is now a function
of 10 parameters $(\phi,\psi,\lambda)$. Again applying the chain rule, we
write 
\begin{eqnarray} 
 \Lambda(\phi,\psi,\lambda| D_{a}, D_{s})  =  
                                    Pr(\phi|D_{a},D_{s}) \:\times &  \\ 
                            \: Pr(\psi|\phi,D_{a},D_{s}) \:\times &\!Pr(\lambda
|\phi,\psi,D_{s}) \nonumber
\end{eqnarray}
where
\begin{equation} 
  Pr(\phi|D_{a},D_{s}) =  \int \Lambda \: d\psi d\lambda
\end{equation}
and
\begin{equation} 
  Pr(\psi|\phi,D_{a},D_{s}) =  \int \Lambda \: d\lambda \:
                                         / \int \Lambda \: d\psi d\lambda 
\end{equation}
Here $Pr(\phi|D_{a},D_{s})$ is the projection of the 10-D posterior distribution
$\Lambda(\phi,\psi,\lambda|D_{a},D_{s})$ onto the 3-D $\phi$-space. The
product $Pr(\psi|\phi,D_{a},D_{s}) \times Pr(\lambda| \phi, \psi, D_{s})$ then 
specifies
how this summed probability is to be distributed first into $\psi$-space
and then into $\lambda$-space.

The dependence of these probability factors on $D_{a}$ and $D_{s}$ merits
comment. 

Both data sets contain information on $\phi = (\log P,e,\tau)$.
Accordingly, $Pr(\phi)$ depends on both $D_{a}$ and $D_{s}$. 

The $\psi$-vector $(A,B,F,G)$ determines the Campbell elements 
$(a,\rm i,\omega,\Omega)$ and vice versa. Since $\omega$ is a 
spectroscopic as well as an astrometric element, $Pr(\psi|\phi)$ must depend on
$D_{s}$ as well as on $D_{a}$.

If $\phi$ and $\psi$ are given, then, since $\omega = \omega (\psi)$, 
the spectroscopic elements $P,e,\tau,\omega$ are known. 
The data $D_{s}$ then suffices to determines the remaining spectroscopic 
elements $\lambda = (\gamma,K_{1},K_{2})$. Thus $Pr(\lambda| \phi, \psi)$ does 
not depend on $D_{a}$.
\section{Likelihoods}
The probability factors defined in Sect.3 are now evaluated using Bayes'
theorem and the appropriate likelihoods. Throughout this paper, we assume
weak, non-informative priors whose impact on posterior 
distributions can be neglected.
\subsection{Astrometry only}
In this case, the posterior distribution is
\begin{equation} 
  \Lambda(\phi,\psi|D_{a}) \propto {\cal L}_{a} 
\end{equation}
where, ignoring a constant factor,
\begin{equation} 
  {\cal L}_{a} = \exp( -\frac{1}{2} \chi^{2}_{a}) 
\end{equation}
and
\begin{equation}
  \chi^{2}_{a} = \Sigma_{n} w_{n}^{a} (x_{n}-\widetilde{x}_{n})^{2}
   +\Sigma_{n} w_{n}^{a} (y_{n}-\widetilde{y}_{n})^{2}
\end{equation}
Because of linearity, $\widehat{\psi}(\phi)$, the minimum-$\chi^{2}$ 
Thiele-Innes 
vector at given $\phi$, is obtained without iteration, and we can write
\begin{equation} 
 \chi^{2}_{a}(\psi|\phi) = \widehat{\chi}^{2}_{a}(\widehat{\psi}|\phi) + 
         \delta \chi^{2}_{a}(\delta \psi|\phi)
\end{equation}
where  $\delta \chi^{2}_{a}$ is the positive increment in $\chi^{2}_{a}$
due to the displacement to $\psi = \widehat{\psi} + \delta \psi$.

Correspondingly, we write
\begin{equation} 
  {\cal L}_{a}(\phi,\psi) = \widehat{{\cal L}}_{a}(\phi) \:  
                                   \widetilde{{\cal L}}_{a}(\psi|\phi)
\end{equation}
where
\begin{equation} 
  \widehat{{\cal L}_{a}} = \exp(- \frac{1}{2} \widehat{\chi}^{2}_{a}) \;\;\; and \;\;\; 
  \widetilde{{\cal L}_{a}} =  \exp( -\frac{1}{2} \delta \chi^{2}_{a}) 
\end{equation}

The statistics of displacements in $\psi$-space is treated in 
Appendix A of L14b. These follow a quadrivariate normal distribution
such that
\begin{equation} 
  Pr(\psi| \phi, D_{a}) = {\cal C}^{-1}   \exp(-\frac{1}{2} \delta \chi^{2}_{a})
\end{equation}
where ${\cal C}(\phi) = (2 \pi)^{2} \sqrt{\Delta}$ and $\Delta$ is 
the determinant of the
covariance matrix. It follows that 
\begin{equation} 
     \widetilde{ {\cal L}_{a} } = {\cal C} (\phi) Pr(\psi| \phi, D_{a})
\end{equation}
Substituting $\Lambda \propto \widehat{{\cal L}}_{a}  \widetilde{{\cal L}}_{a}$
into Eq.(8) and eliminating  $\widetilde{{\cal L}}_{a}$ with Eq.(19),
we obtain
\begin{equation} 
   Pr(\phi|D_{a}) \propto {\cal C}(\phi) \exp(-\frac{1}{2} \widehat{\chi}^{2}_{a})
\end{equation}
This determines the relative weights of the grid points $\phi_{ijk}$ and
agrees with Eq.(14) in L14b.

From a random sampling of the quadrivariate normal distribution 
$Pr(\psi| \phi, D_{a}) $,  we obtain the approximation 
\begin{equation} 
     Pr(\psi|\phi,D_{a}) = {\cal N}^{-1}_{\psi} \sum_{\ell} \delta(\psi - \psi_{\ell})
\end{equation}
Accordingly, the relative weights from Eq.(20) are distributed equally among
the points $\psi_{\ell}$ in $\psi$-space. (Note that at each
$\phi_{ijk}$ an independent sample $\{\psi_{\ell}\}$ is drawn.) 

If the errors in $\widetilde{x}_{n}$ and
$\widetilde{y}_{n}$ are uncorrelated, the quadrivariate distribution 
$Pr(\psi|\phi,D_{a})$ simplifies to the product
of two bivariate normal distributions - see Appendix A in L14b.
\subsection{Astrometry and spectroscopy}
With the addition of spectroscopic data and again assuming
non-informative priors, the posterior density is
\begin{equation} 
 \Lambda(\phi,\psi,\lambda|D_{a},D_{s})  \propto {\cal L}_{a} {\cal L}_{s}
\end{equation}
where, ignoring a constant factor,
\begin{equation} 
  {\cal L}_{s} = \exp( -\frac{1}{2} \chi^{2}_{s}) 
\end{equation}
and
\begin{equation}
  \chi^{2}_{s} = \Sigma_{n} w_{1n}^{s} (v_{1n}-\widetilde{v}_{1n})^{2}
   +\Sigma_{n} w_{2n}^{s} (v_{2n}-\widetilde{v}_{2n})^{2}
\end{equation}
Because of linearity in $\lambda = (\gamma, K_{1},K_{2})$ when $\phi$ and
$\psi$ are fixed, $\widehat{\lambda}(\phi,\psi|D_{s})$, the minimum-$\chi^{2}$
vector, is obtained without iteration, and we can write 
\begin{equation} 
 \chi^{2}_{s}(\lambda|\phi,\psi) = \widehat{\chi}^{2}_{s}(\widehat{\lambda}|\phi,\psi) 
                    + \delta \chi^{2}_{s}(\delta \lambda| \phi,\psi)
\end{equation}
where  $\delta \chi^{2}_{s}$ is the positive increment in $\chi^{2}_{s}$
due to the displacement to $\lambda = \widehat{\lambda} + \delta \lambda$.

Correspondingly, we write
\begin{equation} 
  {\cal L}_{s}(\lambda|\phi,\psi) = \widehat{{\cal L}}_{s}(\phi,\psi) \:  
                                   \widetilde{{\cal L}}_{s}(\lambda|\phi,\psi)
\end{equation}
where
\begin{equation} 
  \widehat{{\cal L}_{s}} = \exp(- \frac{1}{2} \widehat{\chi}^{2}_{s}) \;\;\; and \;\;\; 
  \widetilde{{\cal L}_{s}} =  \exp( -\frac{1}{2} \delta \chi^{2}_{s}) 
\end{equation}

The statistics of displacements in $\lambda$-space is treated in 
Appendix A. These follow a trivariate normal distribution 
$Pr(\lambda|\phi,\psi,D_{s})$ such that Eq.(A.1) holds. From Eqs.(27) and
(A.1), we obtain   
\begin{equation} 
    \widetilde{ {\cal L}_{s}} = {\cal D} (\phi,\psi) Pr(\lambda| \phi,\psi, D_{s})
\end{equation}

The statistics of displacements in $\psi$-space is modified by the
spectroscopic data as noted in Sect.3.3, and this is treated in Appendix B.

We now calculate $Pr(\phi|D_{a},D_{s})$.
Substituting $\Lambda \propto 
      \widehat{\cal L}_{a} \widetilde{\cal L}_{a} \widehat{\cal L}_{s} \widetilde{\cal L}_{s}$
into Eq.(10), eliminating $\widetilde{\cal L}_{s}$ with Eq.(28), and intergrating
over $\lambda$, we obtain
\begin{equation} 
   Pr(\phi|D_{a},D_{s}) \: \propto \: \widehat{\cal L}_{a} \int {\cal D} \: 
                        \widetilde{\cal L}_{a} \widehat{\cal L}_{s} \: d\psi        
\end{equation}
We now eliminate $\widetilde{\cal L}_{a}$ using Eq.(19) to obtain
\begin{equation} 
   Pr(\phi|D_{a},D_{s}) \: \propto \: {\cal C} \widehat{\cal L}_{a} 
            \int {\cal D} \: \widehat{\cal L}_{s} Pr(\psi|\phi,D_{a}) \: d\psi
\end{equation}
If we now replace $Pr(\psi|\phi,D_{a})$ by the approximation given in Eq.(21)
and assume that ${\cal N}_{\psi}$ is independent of $\phi$, then
\begin{equation} 
   Pr(\phi|D_{a},D_{s}) \: \propto \: {\cal C} \widehat{\cal L}_{a} \:
                 \sum_{\ell} ({\cal D} \widehat{\cal L}_{s})_{\psi_{\ell}}
\end{equation}
Accordingly, the relative weights of points $\phi_{ijk}$ in the $\phi$-grid
are
\begin{equation} 
   \mu_{ijk} =  {\cal C} \widehat{\cal L}_{a} \times
                 \sum_{\ell} ({\cal D} \widehat{\cal L}_{s})_{\psi_{\ell}}
\end{equation}
Here the first factor ${\cal C} \widehat{\cal L}_{a}$ depends only on $D_{a}$.
The dependence on $D_{s}$ is introduced by the second factor: If at a 
given $\phi$, all $\psi_{\ell}$ correspond to poor fits to $D_{s}$, then the 
second factor disfavours that $\phi$.
\section{Numerical results}

The technique developed in Sects.3 and 4 is now applied to
synthetic data $D_{a}$ and $D_{s}$ created as described in Sect.2.3 for the 
model binary defined in Sect.2.2. All calculations use a $100^{3}$ grid
for $\phi$-space, and Monte Carlo sampling with ${\cal N}_{\psi} = 20$
for $\psi$-space and ${\cal N}_{\lambda} = 20$ for $\lambda$-space.  
\subsection{Parameter cloud}
Let $\phi_{ijk}$ denote cell mid-points of the 3-D grid spanning $\phi$-space.
At each $\phi_{ijk}$, the technique generates ${\cal N}_{\psi}$ points 
$\psi_{\ell}$ in ${\psi}$-space. Then, at each $\psi_{\ell}$, the technique 
generates
${\cal N}_{\lambda}$ points $\lambda_{m}$ in $\lambda$-space. Thus, 
with this cascade, 
a cloud of points $(\phi_{ijk}, \psi_{\ell}, \lambda_{m})$ is generated in the
10-D $(\phi,\psi,\lambda)$-space. 

Note that this is a cloud of orbit parameters and {\em not} a cloud of orbits.
Because linearity in $\psi$ and $\lambda$ is fully exploited, 
the values of $\chi^{2}_{a}$ and $\chi^{2}_{s}$ at cloud points are derived
without computing astrometric and spectroscopic orbits, and this
is the origin of the technique's computational efficiency.

The relative weights of cloud points $(\phi_{ijk}, \psi_{\ell}, \lambda_{m})$
are
\begin{equation} 
  \mu_{ijk,\ell,m} =  \mu_{ijk} \times \zeta_{\ell} 
                                            \times {\cal N}^{-1}_{\lambda} 
\end{equation}
The first factor comes from Eq.(32), the second from Eq.(B.5),
and the third from Eq.(A.4).    

Note that the third factor is only relevant if ${\cal N}_{\lambda}$
varies with $(\phi,\psi)$.
Note also that if $Pr(\psi|\phi,D_{a},D_{s})$ given by Eq.(B2) were randomly 
sampled, the second factor would be ${\cal N}^{-1}_{\psi}$. Instead,
the quadrivariate normal distribution $Pr(\psi|\phi,D_{a})$ is
sampled and then corrected via the coefficients $\zeta_{\ell}$, which are
such that $\sum \zeta_{\ell} = 1$ - see Appendix B.
\subsection{Inferences}
Suppose $Q(\Theta)$ is a quantity of interest. Its posterior distribution
derived from the parameter cloud is 
\begin{equation} 
    \Theta(Q)  = \sum_{t}  \: \mu_{t} \: \delta(Q-Q_{t}) \:/ 
                                      \: \sum_{t}  \: \mu_{t}   
\end{equation}
where $t \equiv (ijk,\ell,m)$.      
The corresponding cumulative distribution function is
\begin{equation} 
   F(Q)  = \sum_{Q_{t} < Q}  \: \mu_{t} \:/ \: \sum_{t}  \: \mu_{t}   
\end{equation}
The {\em equal tail} credibility interval $(Q_{L},Q_{U})$
corresponding to $\pm 1\sigma$ is then obtained from the equations
\begin{equation} 
 F(Q_{L})  = 0.1587   \;\;\;\;\;     F(Q_{U})  = 0.8413   
\end{equation}
so that the enclosed probability is 0.6826.

These credibility intervals are asymptotically rigorous - i.e., are exact
in the limits ${\cal N}_{\psi} \rightarrow \infty$,
${\cal N}_{\lambda} \rightarrow \infty$, and grid steps $\rightarrow 0$. 
\subsection{An example}  
The fundamental data derivable from the combined astrometric and spectroscopic
data are the component masses ${\cal M}_{1},{\cal M}_{2}$ and the 
parallax $\varpi$. None of these quantities
can be derived if only one data set is available. 

At every cloud point $t$, Kepler's law and the two spectroscopic mass functions
can be solved for ${\cal M}_{1},{\cal M}_{2}$ and $\varpi$, and these values
each have relative weight $\mu_{t}$. Accordingly, the posterior densities of
these quantities can be calculated from Eq.(34) and their credibility intervals
from Eq.(35).

For a particular simulation of $D_{a}$ and of $D_{s}$ , the posterior 
densities so derived are plotted in Figs.1 and 2.
Also plotted are the posterior means, the $1-$ and $2-\!\sigma$ credibility 
intervals, as well as the exact values from Sect.2.2. In each case, the
the exact values fall within the $1-\!\sigma$ limits.

\begin{figure}
\vspace{8.2cm}
\includegraphics{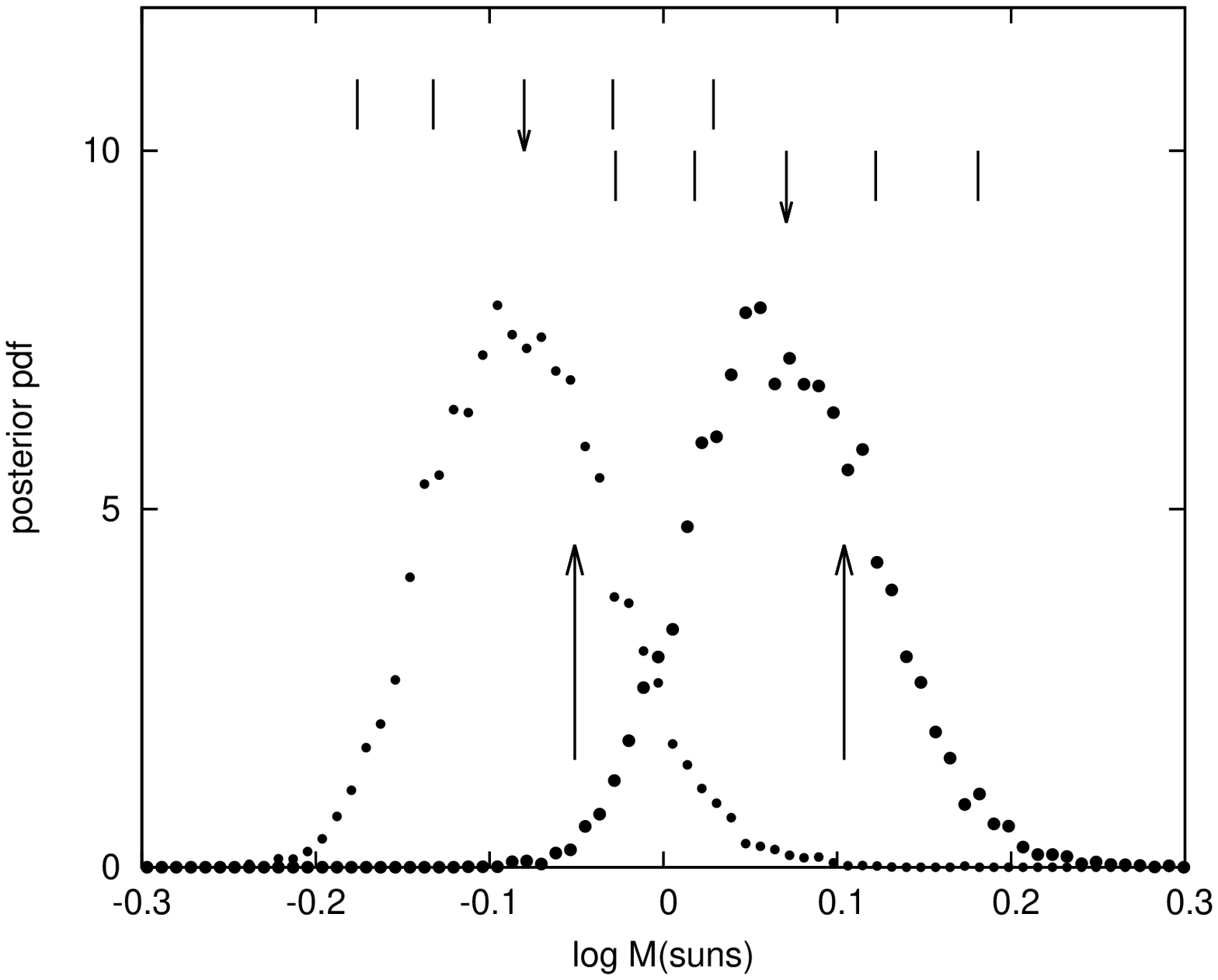}
\caption{Posterior densities for $\log {\cal M}_{1}$ and $\log {\cal M}_{2}$.
The long vertical arrows indicate exact values. The short vertical arrows and 
lines indicate the posterior means and the 1- and 2-$\!\sigma$ credibility
intervals.} 
\end{figure}
\begin{figure}
\vspace{8.2cm}
\includegraphics{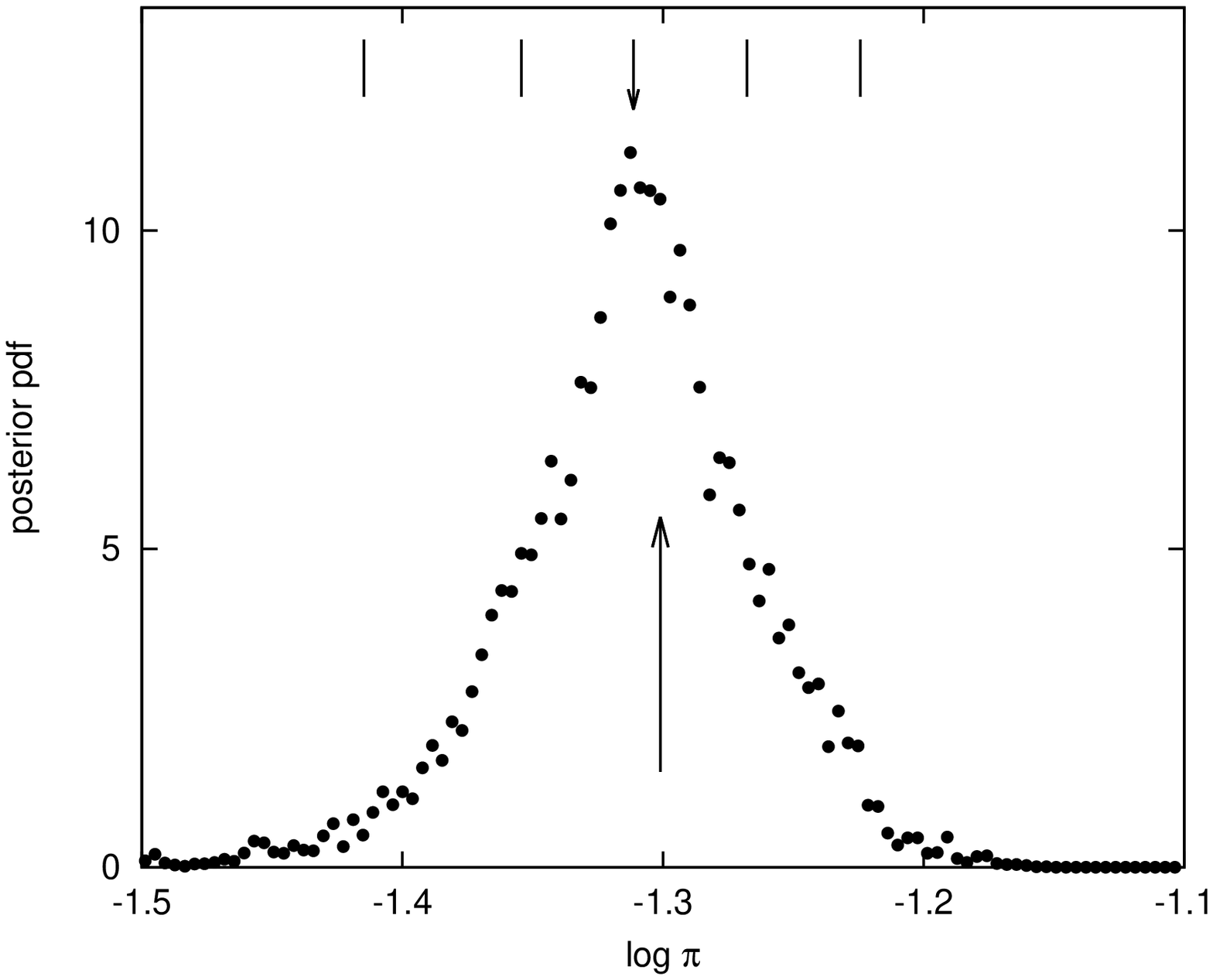}
\caption{Posterior density for $\log \varpi(\arcsec)$.
The long vertical arrow indicates the exact value. The short vertical arrow and 
lines indicate the posterior mean and the 1- and 2-$\!\sigma$ credibility
intervals.} 
\end{figure}

In Appendix C, following L16, a Bayesian goodness-of-fit statistic,
$\chi^{2}_{B}$, is defined
together with corresponding Bayesian $p$-value. We now apply this test. 
For the astrometric data, the posterior mean of $\chi^{2}_{a}$ is
$\langle \chi^{2}_{a} \rangle_{u}$ = 22.2, and for the spectroscopic data
$\langle \chi^{2}_{s} \rangle_{u}$ = 16.9, so that in total
$\langle \chi^{2} \rangle_{u}$ = 39.1. Since the total number of parameters is
$k = 10$, Eq.(C.4) gives $\chi^{2}_{B} = 29.1$.

The total number of measurements is $n = 40$, comprising two astrometric
$(x,y)$ and two spectroscopic $(v_{1},v_{2})$ measurements in each of ten years.
The number of
degrees of freedom is therefore $\nu = n-k = 30$. Substitution in Eq.(C.5)
then gives $p_{B} = 0.51$, a value consistent with the belief that
the data is analysed with a valid model and that Bayesian
inferences are not suspect.
\subsection{Coverage}

An accurate guage of the statistical performance of the technique requires
many repetitions of the above calculation with independently drawn samples
of $D_{a}$ and $D_{s}$.

With different data, the posterior densities and corresponding credibility 
limits in Figs.1 and 2 change. But the long vertical arrows marking
exact values remain fixed. For each independent repetition, we can record
whether or not the exact values are enclosed by the 1- and $2-\!\sigma$
credibility intervals. 
In this way, we carry out a coverage experiment as in L14a,b - see also  
Martinez et al. 2017.

The results obtained from 1000 repetitions are summarized in Table 1.
These show reasonable agreement with $\varepsilon(f)$, the expected fractions
for errors obeying a normal distribution. Thus, despite the non-linearities, 
the credibility intervals retain their conventional interpretations.

\begin{table}

\caption{Coverage fractions from $10^{3}$ trials}

\label{table:1}

\centering

\begin{tabular}{c c c}

\hline
\hline

 $Q$                 &     $1-\!\sigma$       &  $ 2-\!\sigma$      \\

\hline                                                                  
                                                                         \\

 ${\cal E}(f)$     &   $0.683 \pm 0.015$  &    $0.954 \pm 0.007$ \\ 

                                                                         \\

 $ \log {\cal M}_{1} $  &   $0.711 \pm 0.014$  &  $0.964 \pm 0.006$ \\ 

 $ \log {\cal M}_{2}$       &   $0.732 \pm 0.014$  & $ 0.963 \pm 0.006$ \\ 
                                                                     \\
 $ \log \varpi$    &   $0.667 \pm 0.015$  & $ 0.963 \pm 0.006$ \\

\\

\cline{1-3}

\hline
\hline

\end{tabular}

\end{table}

\section{Hypothesis testing}

In L16 and references therein, the relative absence in the astronomical
literature of statistical testing of Bayesian models is commented upon.

\subsection{Some quotes}

The following quote from a statistician (Anscombe 1963)
indicates that concern on this issue is of long standing:\\ 

``To anyone sympathetic with the current neo-Bernoullian neo-Bayesian Ramseyesque Finettist Savageous movement in statistics, the subject of testing goodness of fit is something of an embarrassment.''\\

A very recent comment (Fischer et al. 2016 in Sect.4.1, authored by E.Ford) 
referring to exoplanets is:\\    

``Too often people using Bayesian methods ignore model checking, because it
doesn't have a neat and tidy formal expression in the Bayesian approach.
But it is no less necessary to do goodness-of-fit type checks for a 
Bayesian analysis than it is for a frequentist analysis''\\ 
\subsection{Additional justifications}  
When authors ignore model checking, they seldom, if ever, explain why.
The quote above suggests that Bayesians are deterred by the absence of
a readily-applied test. In contrast, frequentists reporting a
minimum-$\chi^2$ analysis 
generally include $\chi^{2}_{0}$, the $\chi^{2}$ minimum, and often also 
the $p$-value derived from the known distribution of $\chi^{2}_{0}$.
Thus, this traditional, frequentist approach has a {\em built-in} reality 
check. Moreover, this check is rigorously justified for linear models and 
normally-distributed measurement errors.

Note that minimum-$\chi^2$ codes return estimates and confidence intervals
even when $\chi^{2}_{0}$ corresponds to a vanishingly small $p$-value. 
Thus, we may surmise that innumerable spurious inferences from 
false hypotheses or poor data are absent from the scientific literature 
precisely because of this built-in check.

Besides the difficulty of Bayesian model-checking, it seems likely that the 
following additional reasons play a role in checking being ignored:\\  

The detection of the expected signal confirms the hypothesis.\\

This is endemic in studies of orbits, including frequentist
analyses going back decades. If a star is investigated for reflex motion
due to a companion and a periodic signal is detected, then it is all too easy to
take this as confirmation of a companion. A more critical approach recognizes
that a harmonic expansion of Keplerian motion provides  quantitative tests
of the orbit hypothesis. When this approach is applied to a sample of 
spectroscopic binaries with exquisitely accurate radial velocities,
significant departures from exact Keplerian motion are found
(Lucy 2005, Hearnshaw et al. 2012).

A notable recent signal detection is that of gravitational waves from coalescing
black holes (Abbott et al. 2016). The published parameters for the
initial black hole binary derive from a Bayesian analysis. But 
these authors do not ignore model checking: their Bayesian analysis   
is preceded by a standard frequentist $\chi^{2}$ test of template fits.\\

The Bayesian model has so many parameters that a poor fit is improbable.\\

In this case, the acceptance-rejection aspect of the scientific method is
replaced by the posterior density favouring or disfavouring regions of
parameter space. The expectation is that with enough high quality data, the 
posterior density will be sharply peaked at the point corresponding to the 
true hypothesis.
But what if the true hypothesis is not part of
the adopted multi-parameter model? How does the investigator detect this?\\

A hypothesis should not be rejected if there is no alternative.\\ 

On this view, Bayesian model checking, even if readily carried out, should
not lead to the rejection of a hypothesis. Rather, one should wait until
an alternative hypothesis is proposed and then implement the
model selection machinery. This view goes back to Jeffreys (1939, Sect. 7.2.2)
- see also Sivia and Skilling (2006, p.85).

Jeffreys supports this view by remarking that there was never a time over
previous centuries when Newton's theory of gravity would not have failed
a $p$-test. It is therefore instructive to recall how Adams and Le Verrier 
reacted
to the large residuals in the motion of Uranus - i.e., small $p$-value. 
Crucially, their view was that the hypothesis being tested was not 
Newton's theory    
but the then current 7-planet model of the solar system. 
Because they had a far greater
degree of belief in Newtonian gravity than in the 7-planet model, they 
doubted the latter and went on to successfully predict Neptune's position.
This example illustrates that ambitious and effective scientists take small
$p$-values seriously even in the absence of alternative hypotheses.
By doing so, they create alternative hypotheses.

\subsection{The $\chi^{2}_{B}$ statistic}  
A Bayesian goodness-of-fit statistic $\chi^{2}_{B}$ and corresponding 
$p$-value is
defined in Appendix C. 

A crucial requirement of any goodness-of-fit (GOF) 
statistic is that it should not often falsely lead one to reject or doubt an 
hypothesis when that hypothesis is true. In frequentist terms, this   
is a Type II error. Such an error arises when the statistic gives a small
p-value, say p = 0.001, even when the null hypothesis $(H_{0})$ is true. Of 
course, such a value can occur by chance, but for
an acceptable GOF statistic the frequency of Type II errors 
should not markedly exceed $p$.

In the simulation reported in Sects.5.3 and 5.4, the null hypothesis is
correct by construction, since the data is generated from the exact
formulae for the astrometric and spectroscopic orbits. Thus, if the
mathematical models were completely linear, we would expect $\chi^{2}_{B}$
to be distributed exactly as  $\chi^{2}_{\nu}$ with $\nu = n-k$ degrees of
freedom. The $p$-value defined by Eq.(C.5) would then have an exactly 
uniform distribution in the interval $(0,1)$.

The $N_{tot} = 1,000$ simulations used for the coverage experiment in 
Sect.5.4 allow the uniformity of the $p_{B}$ values to be tested. 
In Fig.3, the fraction with $p_{B} < p$ is plotted against $p$. We see
that uniformity is obeyed with reasonable precision for $p \in (0.01,1.00)$.
In particular, there is no indication of a significant departure from
uniformity that could be attributed to the non-linearities. Accordingly, 
the $p_{B}$ values derived from $\chi^{2}_{B}$ can be interpreted in the same 
way and with the same confidence as $p$-values in minimum-$\chi^{2}$ estimation.

Note that the calculation of $\chi^{2}_{B}$ is a trivial addition to an existing
Bayesian code that very likely already calculates the posterior means of other
quantities.

\begin{figure}
\vspace{8.2cm}
\includegraphics{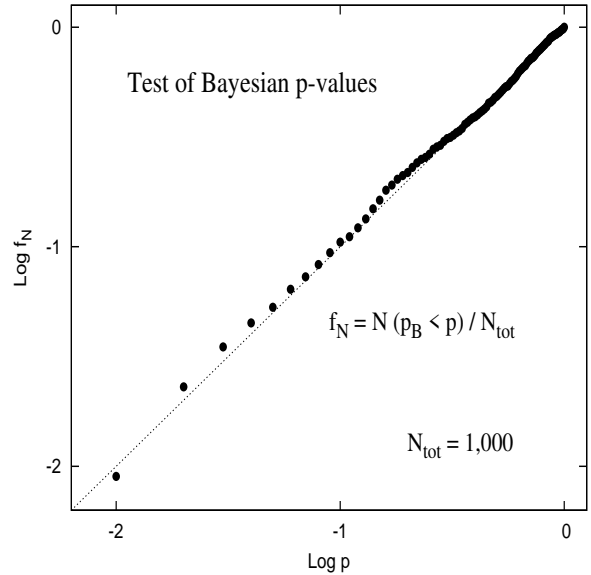}
\caption{Test of Bayesian $p$-values. From 1,000 simulations, the fraction 
with $p_{B} > p$ is plotted against $p$ for 
$p = 0.01(0.01)1.00$. 
The dashed line shows the expected result if the null hypothesis is 
correct {\em and} if the statistic $\chi^{2}_{B}$ obeys the $\chi^{2}_{\nu}$ 
distribution with $\nu = n -k$ degrees of freedom.} 
\end{figure}
\subsection{Posterior predictive $p$-values}  
In the contribution authored by E.Ford from which the quote in Sect.6.1
is taken, readers are referred to Gelman et al.(2013) who
recommend posterior predictive $p$-values for testing Bayesian models.

In the context of the technique developed here, this recommendation proceeds as 
follows: 1) Randomly select a point in parameter space from the posterior
distribution. Thus, if $t$ is an index that gives a 1-D enumeration of
the parameter cloud, a random point $t'$ is that which most closely satisfies
the equation  
\begin{equation} 
       \sum_{t < t'} \mu_{t} \: /\:  \sum_{t} \mu_{t} = x
\end{equation}
where $x$ is a random number $\in (0,1)$.\\ 
2) From the 10 parameters at $t'$, create synthetic data 
$D' = D'_{a} + D'_{s}$, 
compute $\chi^{2} = \chi^{2}_{a} + \chi^{2}_{s}$, and then compare to the   
$\chi^{2}$ at $t'$ for the original data $D = D_{a} + D_{s}$.\\
3) Repeat steps 1) and 2) ${\cal N}_{tot}$ times.

A Bayesian $p$-value is then defined to be
\begin{equation} 
       p_{B} = {\cal N} ( \chi^{2} (D') > \chi^{2} (D) ) \:/\: {\cal N}_{tot} 
\end{equation}
Thus a small value of $p_{B}$ indicates that it is hard to find points $t'$
giving a worse fit than the original data, indicating that original data
gives a poor fit.

To quote E.Ford again (Sect.6.1), posterior predictive checking is evidently
not 'a neat and tidy' formalism. Moreover, physical scientists have a strong
interest in having reliable $p$-values when $p \la 0.001$, since such values
raise serious doubts about a model's validity. This then requires
${\cal N}_{tot} \sim 10,000$ repetitions of the above steps, which may be
infeasible.   

Posterior predictive $p$-values have been compared to the values given by 
Eq.(C.5) for a simple 1-D toy model. Specifically, a Hubble flow in 
Euclidean space
populated by perfect standard candles. Synthetic data is created and the
posterior density for the Hubble constant derived. A poor fit can be 
engineered by corrupting the data at high redshift and then comparing
the resulting two small $p$-values. They agree closely. 

This suggests that the readily
calculated $p_{B}$ given by Eq.(C.5) eliminates any need for the cumbersome 
direct calculation of the posterior predictive $p$-value given by Eq.(38).

\section{Conclusion}
In this paper, a non trivial example of a wide class of problems in 
statistical astronomy is addressed. These are the so-called hybrid problems 
where the mathematical models predicting the observations are partly linear and
partly non-linear in the basic parameters. As in the simpler,
purely astrometric case considered in L14b, when spectroscopic data
is added, a grid
search over the non-linear parameter space combined with Monte Carlo sampling
in the linear parameter spaces still leads to a computationally efficient
scheme and again yields credibility intervals with close to correct coverage
(Sect 5.4), 
a result of prime importance generally, but especially so when estimating 
fundamental stellar parameters.

In contrast to L14a, the formulation in Sects.3 and 4 is mostly quite general
and so should be readily adapted to other hybrid problems. 

In addition to exhibiting correct coverage, the large number of
independent simulations allow the testing (Sect.6.3) of $\chi^{2}_{B}$, 
a Bayesian goodness-of-fit 
criterion (Appendix C) for posterior probability densities. Even though
the test problem involves some non-linear parameters, the exact 
sampling distribution in the linear case is closely followed, thus
providing a readily-calculated $p$-value that quantifies one's
confidence in the inferences drawn from the posterior distribution.
Since in problems that are exactly linear, the Bayesian and frequentist 
$p$-values are identical, investigators can make the decisions on the basis 
of the Bayesian $p_{B}$-value exactly as they would for a frequentist 
$p$-value. Moreover, since
calculating the statistic $\chi^{2}_{B}$ involves trivial changes to a 
Bayesian code, it provides 'the neat and tidy formal expression' that
is missing in current Bayesian methodology - see quote from E.Ford in
Sect.6.1.

\acknowledgement

The issue of error underestimation in hybrid problems 
was raised by the referee of L14a and was the direct
stimulus of L14b and of this investigation. A useful correspondence
with E.L.N.Jensen is also acknowledged.

\appendix

\section{Statistics in $\lambda$- space}

Statistics in the 4-D Thiele-Innes $\psi$-space is treated in Appendix A of 
L14b.
Analogous results are briefly stated here for the 3-D $\lambda$-space.

Given $\phi$ and $\psi$, the minimum-$\chi^{2}_{s}$ vector 
$\widehat{\lambda} = (\widehat{\gamma} , \widehat{K_{1}}, \widehat{K_{2}})$ 
is obtained without iteration
from the normal equations derived from Eqns. (1) and (24).

The displacement $\lambda = \widehat{\lambda} + \delta \lambda$ gives
$\chi^{2}_{s} = \widehat{\chi}^{2}_{s} + \delta \chi^{2}_{s}$ with positive
$\delta \chi^{2}_{s}$. On the assumption of normally-distributed
errors, the probability density at $\lambda$ is a 
trivariate normal distribution such that
\begin{equation} 
 Pr(\lambda|\phi,\psi,D_{s}) = {\cal D}^{-1} \exp(-\frac{1}{2} 
                                                  \delta \chi^{2}_{s})  
\end{equation}
where ${\cal D}(\phi,\psi) = (2 \pi)^{3/2} \sqrt{\Delta}$ and $\Delta$ is 
the determinant of the covariance matrix.
\subsection{Random sampling in $\lambda$-space}
Points $\lambda_{\ell}$ randomly sampling the trivariate normal distribution
$Pr(\lambda|\phi,\psi,D_{s})$ are derived with a standard procedure 
(Gentle 2009) for sampling multivariate distributions. The first step is to
make a Cholesky decomposition (Press et al.2007,p.100) of the covariance matrix 
$\vec{C}$ - i.e., to find the lower triangular matrix $\vec{L}$ such that
\begin{equation} 
 \vec{L} \vec{L}^{'} = \vec{C} 
\end{equation}
A random sample from $Pr(\lambda|\phi,\psi,D_{s})$ is then
\begin{equation} 
  \lambda = \widehat{\lambda} + \vec{L}.\vec{z} = 
                            \widehat{\lambda} + \delta{\lambda} 
\end{equation}
where the elements of $\vec{z} = (z_{1},z_{2},z_{3})$ are random Gaussian
variates. The resulting approximation to $Pr(\lambda)$ is 
\begin{equation} 
      Pr(\lambda|\phi,\psi,D_{s}) = {\cal N}^{-1}_{\lambda} \sum_{m}
                                         \delta(\lambda - \lambda_{m})
\end{equation}
Note that ${\cal N}_{\lambda}$ can vary with $(\phi,\psi)$. 
The increment in $\chi^{2}_{s}$ due to the displacement
from $\widehat{\lambda}$ is
\begin{equation} 
  \delta \chi^{2}_{s} = z_{1}^{2}+z_{2}^{2}+z_{3}^{2} 
\end{equation}
Accordingly, as in the analogous problem in $\psi$-space - see Eq.(A.22)
in L14b, the increment in $\chi^{2}$ is obtained without computing the
spectroscopic orbits at $\widehat{\lambda} + \delta{\lambda}$ - though this
should be checked during code development. This is a
consequence of linearity and accounts for the computational efficiency
of the technique.   

In Appendix A of L14, Cholesky decompostion is not needed because
the quadrivariate normal distribution $Pr(\psi|\phi,D_{a})$ is the product
of bivariate distributions. But this simplification is lost if 
$\widetilde{x}_{n}$ and $\widetilde{y}_{n}$ have correlated errors (Sect.2.3). 
In that
circumstance, the above Cholesky approach is the necessary generalization.  
\section{Modified statistics in $\psi$- space}
The treatment of statistics in $\psi$-space in Appendix A of L14b does
not apply when spectroscopic data is included. As noted in Sect.3.3 - see
Eq.(11), $Pr(\psi)$ depends on both $D_{a}$ and $D_{s}$ 

The required modification is obtained by substituting
$\Lambda \propto 
      \widehat{\cal L}_{a} \widetilde{\cal L}_{a} \widehat{\cal L}_{s} \widetilde{\cal L}_{s}$
into Eq.(11), integrating over $\lambda$ using Eq.(28), and noting that
$\widehat{\cal L}_{a}$ is independent of $\psi$. This gives
\begin{equation} 
   Pr(\psi|\phi,D_{a},D_{s}) \: \propto \: {\cal D} \: 
                                  \widetilde{\cal L}_{a} \widehat{\cal L}_{s}
\end{equation}
We now eliminate $\widetilde{\cal L}_{a}$ using Eq.(19) and noting that
${\cal C}$ is independent of $\psi$. This gives
\begin{equation} 
   Pr(\psi|\phi,D_{a},D_{s}) \: \propto \: {\cal D} \: 
                                \widehat{\cal L}_{s} \: Pr(\psi|\phi,D_{a})
\end{equation}
showing that $Pr(\psi)$ is modified from the pure astrometry case by the
factor ${\cal D}\widehat{\cal L}_{s}$ introduced by spectroscopy. Because 
of this modification, $Pr(\psi|\phi,D_{a},D_{s})$ is not a multivariate normal
distribution and so not as readily sampled.

The adopted sampling procedure is as follows: from Eq.(A.20) in L14b, we have
the approximation 
\begin{equation} 
   Pr(\psi|\phi,D_{a}) = {\cal N}^{-1}_{\psi} \sum_{\ell} 
                                      \delta( \psi - \psi_{\ell})  
\end{equation}
where the $\psi_{\ell}$ randomly sample the quadrivariate normal distribtion
appropriate in the pure astrometry case (Appendix A, L14b). Substituting 
into Eq.(B.2), we obtain the corresponding approximation when spectroscopy is
included
\begin{equation} 
   Pr(\psi|\phi,D_{a},D_{s}) \: = \:\sum_{\ell} \zeta_{\ell} \: 
                                      \delta( \psi - \psi_{\ell})  
\end{equation}
where
\begin{equation} 
   \zeta_{\ell}(\phi) = ({\cal D} \widehat{\cal L}_{s})_{\psi_{\ell}} \: / \:
                       \sum_{\ell}   ({\cal D} \widehat{\cal L}_{s})_{\psi_{\ell}}
\end{equation}
As might be expected, because of the factor $(\widehat{\cal L}_{s})_{\psi_{\ell}}$,
points $\psi_{\ell}$ in $\psi$-space 
are strongly disfavoured if that $\psi_{\ell}$ gives a poor fit to the 
spectroscopic data. 
\section{The $\chi^{2}_{B}$ and $\psi^{2}$ statistics}  
In an earlier paper (Lucy 2016; L16), we define  
\begin{equation} 
   \langle \chi^{2} \rangle_{u} \:  = \int \chi^{2}(\alpha) 
                                 \Lambda(\alpha|D) \: dV_{\alpha}
\end{equation}
where 
\begin{equation} 
   \Lambda(\alpha|D) \: =  {\cal L} (\alpha|D)  \:/ 
                        \int  {\cal L} (\alpha|D) \:  dV_{\alpha}
\end{equation}
Thus  $\langle \chi^{2} \rangle_{u}$ is the posterior mean of 
$\chi^{2}(\alpha)$ when the posterior density $\Lambda$ is computed
under the assumption of a uniform ($u$) prior.

If the model is linear in the parameter vector $\alpha$ and if errors
are normally-distributed, then (Appendix A, L16)
\begin{equation} 
   \langle \chi^{2} \rangle_{u} \:  = \chi^{2}_{0} + k
\end{equation}
where $\chi^{2}_{0}$ is the minimum value of $\chi^{2}(\alpha)$ and $k$ is
the number of parameters. Moreover, under the stated assumptions,
$\chi^{2}_{0}$ is distributed as $\chi^{2}_{\nu}$, where $\nu = n - k$
is the number degrees of freedom and $n$ is the number of measurements.

It follows that if we define the statistic
\begin{equation} 
   \chi^{2}_{B} =  \langle \chi^{2} \rangle_{u} \: - k
\end{equation}
then, for a linear model and normally-distributed errors, $\chi^{2}_{B}$
is distributed as $\chi^{2}_{\nu}$ with $\nu = n - k$. Accordingly,
a $p$-value that quantifies the quality of the posterior distribution
$\Lambda(\alpha|D)$ from which all Bayesian inferences are drawn
is given by
\begin{equation} 
   p_{B} =  Pr(\chi^{2}_{\nu} > \chi^{2}_{B})  \;\;  for \;\;  \nu = n-k
\end{equation}
If the model is indeed linear in $\alpha$, it follows from Eqs (C.3)
and (C.4) that $\chi^{2}_{B} = \chi^{2}_{0}$, and so the frequentist and
Bayesian $p$-values agree, a gratifying result.

If the model is non-linear in some parameters, then this GOF test should
still be useful if the data is such that the fractional error of the
non-linear parameters are small, for then a linearized model could be used.

In most Bayesian analyses in astronomy, the imposed priors are uninformative
and so this analysis holds. In the rare cases where an informative prior $\pi$
is imposed, perhaps from a previous experiment, the discussion in L16, 
Sect.4.1 suggests that the criterion $\chi^{2}_{B}$ with
$\langle \chi^{2} \rangle_{\pi}$ replacing  
$\langle \chi^{2} \rangle_{u}$ will have closely similar characteristics.

\subsection{Generalization}  

The above analysis assumes uncorrelated measurement errors. The inclusion of
correlations is a straightforward application of the statistics of
quadratic forms - see, e.g., Hamilton (1964, Chap.4). 

When corellations
are included, the $\chi^{2}$ summation is replaced by
\begin{equation} 
   \psi^{2} =  \vec{v}' \vec{M}^{-1} \vec{v}
\end{equation}
where $\vec{M}$ is the covariance matrix and $\vec{v}$ is the vector of
residuals. The previous analysis assumes that the off-diagonal elements
of $\vec{M}^{-1}$ are zero.

With linearity in the parameter vector $\vec{\alpha}$ and normally-
distributed measurement errors $\tilde{\vec{x}}-\vec{x}$, the sampling
distribution of $\tilde{\vec{x}}$ is a $k$-dimensional multivariate normal
disribution $\propto \exp(-\psi^{2}/2)$, where $k$ is the number of parameters.
It follows that the likelihood is
\begin{equation} 
  {\cal L}(\vec{\alpha}|D) \propto  \exp( - \frac{1}{2} \psi^{2})
\end{equation}
Exploiting linearity in $\vec{\alpha}$ and assuming a weak prior, we can
write the posterior density as
\begin{equation} 
  \Lambda(\vec{\alpha}|D) \propto  \exp( - \frac{1}{2} \psi^{2}_{0})
                 \times    \exp( - \frac{1}{2} \delta \psi^{2})
\end{equation}
where $\psi^{2}_{0}$ is the minimum of $\psi^{2}$ at $\vec{\alpha}_{0}$ and
$\delta \psi^{2}$ is the positive increment due to the displacement 
$\vec{\alpha} - \vec{\alpha}_{0}$.
      
Accordingly, in the case of a uniform prior, the posterior mean of $\psi^{2}$
is
\begin{equation} 
  \langle{\psi^{2}}\rangle_{u} =  \psi^{2}_{0} +
        \frac{ \int  \Delta \psi^{2} \exp( -\Delta \psi^{2} /2) \: dV_{\alpha} }
          {\int \exp( -\Delta \psi^{2} /2) \: dV_{\alpha} }
\end{equation}
This has the same form as Eq.(A.4) in L16, with $\Delta \psi^{2}$ replacing
$\Delta \chi^{2}$. Therefore, since surfaces of constant $\Delta \psi^{2}$ are
also self-similar $k$-dimensional ellipsoids, we immediately have
\begin{equation} 
  \langle{\psi^{2}}\rangle_{u} =  \psi^{2}_{0} + k
\end{equation}

Now $\psi^{2}_{0}$ is distributed as $\chi^{2}_{\nu}$ with $\nu = n-k$ degrees of
freedom. Accordingly, the statistic 
\begin{equation} 
  \psi^{2}_{B} =  \langle{\psi^{2}}\rangle_{u} - k
\end{equation}
is distributed as $\chi^{2}_{\nu}$ with $\nu = n-k$ degrees of
freedom.  

This is the required generalization of $\chi^{2}_{B}$ given by Eq.(C.4).

\end{document}